# Competitive electrohydrodynamic and electro-solutal advection arrests evaporation kinetics of droplets

Vivek Jaiswal[1], Shubham Singh[1], A R Harikrishnan[2] and Purbarun Dhar[1, *]

[1]Department of Mechanical Engineering, Indian Institute of Technology Ropar, Rupnagar–140001, India

[2]Department of Mechanical Engineering, Indian Institute of Technology Madras, Chennai–600036, India

* *Corresponding author*:

E–mail: purbarun@iitrpr.ac.in

Phone: +91–1881– 24–2173

**Abstract**

The present article reports the hitherto unreported phenomenon of arrested evaporation dynamics in pendent droplets in an electric field ambience. The evaporation kinetics of pendant droplets of electrically conducting saline solutions in the presence of a transverse, alternating electric field is investigated experimentally. It has been observed that while increase of field strength arrests the evaporation, increment in field frequency has the opposite effect. The same has been explained on the solvation kinetics of the ions in the polar water. Theoretical analysis reveals that change in surface tension and diffusion driven evaporation model cannot predict the arrested or decelerated evaporation. With the aid of Particle Image Velocimetry, suppression of internal circulation velocity within the droplet is observed under electric field stimulus, and this affects the evaporation rate directly. A mathematical scaling model is proposed to quantify the effects of electrohydrodynamic circulation, electro-thermal and electro-solutal advection on the evaporation kinetics of the droplet. The analysis encompasses major governing parameters, viz. the thermal and solutal Marangoni numbers, the Electrohydrodynamic number, the electro-Prandtl and electro-Schmidt numbers and their respective contributions. It has been shown that the electro-thermal Marangoni effect is supressed by the electric field, leading to deteriorated evaporation rates. Additionally, the electro-solutal Marangoni effect further supresses the internal advection, which again arrests the evaporation rate by a larger proportion. Stability analysis reveals that the electric body force retards the stable internal circulation within such droplets and arrests advection. The stability mapping also illustrates that if the field strength is high enough for the electro-solutal advection to overshadow the solutal Marangoni effect completely, it can lead to improvement in evaporation rates. The present findings on the competitive electrohydrodynamic and electro-solutal behavior within pendant droplets may find strong implications in microscale flows involving aspects such as droplet electrokinetics, dielectrophoresis, and electrostriction and so on.

# 1. Introduction

Understanding the fluid dynamics, thermal and species transport behaviour in microscale droplets and/or sprays has garnered research interest for quite some time due to continuous evolution and advancement of systems using droplets as the working entity. The recent research and development activities in the area of droplet dynamics have been triggered by the range of applications where droplets act as the dominant component. The major applications are in bio-medical devices (spray medications, DNA patterning, pathological test (blood culture and diabetics profiling), nebulizer and inhaler based drug delivery sprays, etc.), automobiles and internal combustion engines [1-3], manufacturing and thermal treatment [4], printing technology, and the HVAC sector. The bio-medical application includes. The combustion engine sector requires understanding of droplet and spray dynamics for design of combustion chamber, ignition systems and fuel injectors. Insecticides, fertilizer sprays and repellents [5-6] and microfluidic devices such as lab-on-chip devices and droplet based sorting fluidic devices are other applications where droplet dynamics play important roles. Consequently, finding novel phenomena in the fluid dynamics, thermal as well as the mass transfer in droplets is at the crux of engineering existing as well as new possibilities of application.

There are two types of droplets generally encountered as sessile and pendant. The sessile droplet sits on the substrate and takes suitable shapes respective to surface morphology. As the name suggests, droplet hanging from solid surface takes the shape of the pendant and hence called pendant droplet. Other properties such as thermophysical properties: density, surface tension, and surface energy are also involved in the kinetics of sessile droplets. A major focus of past studies has been in the area of sessile droplet where concentration was on interfacial properties to be in the center of change in evaporation dynamics of the droplet. Authors like Picknett and Bexon [7], Hu and Larsen [4], Fukai et al. [8], and Semenov et al. [9], etc. are some to be named in the field of sessile droplet evaporation. Studies related to triple line analysis, evaporative flux at droplet interface and effect of surface wettability and substrate temperature variation defines diversity in the effort. But in the recent times, researchers turned towards pendant droplet for an interesting case of non-reliance on interfacial phenomena due to very less surface interaction. As of record states, first evolutionary work mentioned in literature was done by Godsave [10] where the assumption of evaporation in the quasi-static gas phase of a spherical droplet with constant diffusion rate. The droplet temperature and pressure were kept constant. With above assumptions, the $D^2$ law was proposed which is used to date for compliance with stated theory. A model developed by Kuz [11] which comprises Hertz- Knudsen criteria as well as follows the $D^2$ law. The Hertz- Knudsen criteria are used to relate pressure with time and concentration rate while evaporation happens. Some of the recent development in pendant droplet studies are involvement of multi-physics such magnetic field (Laroze [12]). The complexity of different physics involvement gives rise to internal advection effect due to concentration change as well as due to misalignment in the orientation of particle pole to the field setting. There are characteristics changes in flow behavior as well as transfer mechanism as reported by Rossow [13] and Jaiswal et al. [14]. Some other multi-physics studies still in pipelines with different heat and mass transfer mechanism.

As an application of multi-physics is discussed, electric field also poses as a potential candidate as a research center in droplet evaporation arena. To overcome the limitation of the only diffusion driven mechanism, only a few works have been done to date in the area of droplet evaporation under the influence of electric field. The shape of the droplet is also governed by surface tension, gravity, and in this case on the electric field also. All other interfacial characteristics also vary due to the interception of field effect with gravity and surface tension.

Studies related to determination or tracking droplet shape under electrostatic force can list Bateni et al. [15-17], Reznik et al. [18], Basaran and Wohlhuter [19] and Miksis [20] as a remarkable contribution. They all numerically targeted the deviation of the axisymmetric droplet with the orientation of the electric field. New numerical tools such as ADSA-EF (axisymmetric drop shape analysis- electric field) were developed for this purpose, and their results fairly match with experimentation. Combination of Laplace, Young- Laplace, Maxwell and Stokes equations together form the algorithm to devise ADSA-EF technique. In the case of the sessile droplet, the balance was reported as augmented Young- Laplace equation (Adamiak [21]) where surface energy equated to the summation of pressure energy, potential energy, and electric field energy. The electric field component is expressed regarding electrical pressure difference which difference in normal Maxwell stress tensor across the interface of the droplet. The electric field environment exerts a volumetric electric force which combination of Coulomb force, dielectrophoretic force, and electrostriction term. The Coulomb force arises due to free charge particle interaction inside the fluid whereas dielectrophoretic (DEP), and electrostriction force constitutes of non- uniform electric field and non - uniform dielectric permittivity. The DEP force depends on the frequency of electric field, medium as well as particle electrical properties. Since all biological cells have dielectric properties, DEP can be used to devise bio-medical application such as platelet sorter, carcinogenic tissue separation, drug discovery mechanism, cell therapeutics and biological particle sorter designing. Unlike to it, electrostriction force arises due to the dissimilarity of dielectric permittivity which causes strain propagation throughout the bulk material. The strain is resultant of alignment shift by cation and anion of dielectric material with the onset of the electric field.

A thorough literature survey gives certain takeaway from the field of electric field influenced pendant droplet evaporation. The kinetics and representation of evaporation dynamics under field influence is an untouched arena. The majority of studies concentrate on computationally deriving the effect of electric field on droplet evaporation and estimating shape taken during evaporation. Therefore, the study of the effect of the electric field stimuli on heat and mass transport phenomena through experimental approach lacks and necessary to set the understanding benchmark for manipulation of droplet evaporation. The application like DEP requires more experimental support to check the extent suggested by computational domain of implementation. The present paper focusses on understanding behavior and conjugation of fluid and heat dynamics in evaporation of aqueous saline solution with the help of experimentation under electric field stimuli. The presence of ionic inclusion is already explored arena and effect of concentration on the change in evaporation pattern of pendant droplet is discussed the topic (Jaiswal et al. [14]). The rise in the concentration of solute poses positive effect on evaporation of droplet and evaporation characteristics enhancement is also proportional to the solubility of the dissolved salt. The augmentation was supported by the presence of advance internal circulation which was confirmed by PIV (particle image velocimetry). The solutal advection was reported as a cause of enhancement of internal circulation. In the presence of electric field, the ionic configuration of solute also matters as it will orient/ align according to electric field set-up. This orientation shift can be achieved through variation voltage as well as frequency. The modulation of evaporation dynamics of saline pendant droplet through a change in voltage and frequency is in center of study in the present paper. The understanding of above topic will find application in many thermo-fluidic devices, bio-medical as well as aeronautical application such as DEP.

## 2. Materials and methodologies

A customized experimental setup has been used for the experiments and the schematic of the setup has been illustrated in Fig. 1. The droplet is dispensed from a steel needle using a digitally controlled precision dispensing mechanism (Holmarc Opto-mechatronics Pvt. Ltd., India). A glass syringe of 50 ± 0.1 µL capacity has been used and a droplet of 20 ± 0.5 µL volume was suspended as a pendant from a stainless steel needle. The nominal diameter of the droplet lies in the range of 2.9 ± 0.2 mm. Two electrodes have been arranged for the purpose of applying the electric field across the droplet in a vertical orientation. The field was generated by the use of a programmable AC power supply (Aplab, India) with caliber of 50 to 500 Hz at 0.1 Hz resolution and 0-270 V at 0.1 V resolution. One terminal of the supply was connected to the steel needle, which acted as one electrode. The other terminal was connected to a steel plate electrode suspended beneath the hanging droplet (at a distance of 1 mm from the droplet bottom tip) . This vertical arrangement of field application has been illustrated in the inset of figure 1. The total distance between needle electrode and the plate electrode was maintained at 4 ± 0.2 mm. Four electric field strengths, , viz. 100, 150, 200 and 250 V and three frequency settings, viz. 50, 100 and 200 Hz, have been used. The control case for pure water was compared to literature reports [14, 22] to validate the experimental setup. The pure water is also tested under the influence of variant frequencies and field strengths and insignificant deviation was observed from the zero-field case. Two salts, NaI (sodium iodide) and $CuSO_4.5H_2O$ (copper pentahydrate) were used obtain salt solution droplets. Both the salts were procured from Sigma Aldrich, India and were used as obtained and the salts were chosen based upon findings reported by the present authors [14]. Four concentrations, viz. 0.01, 0.05, 0.1 and 0.2 M were used for the present experiments.

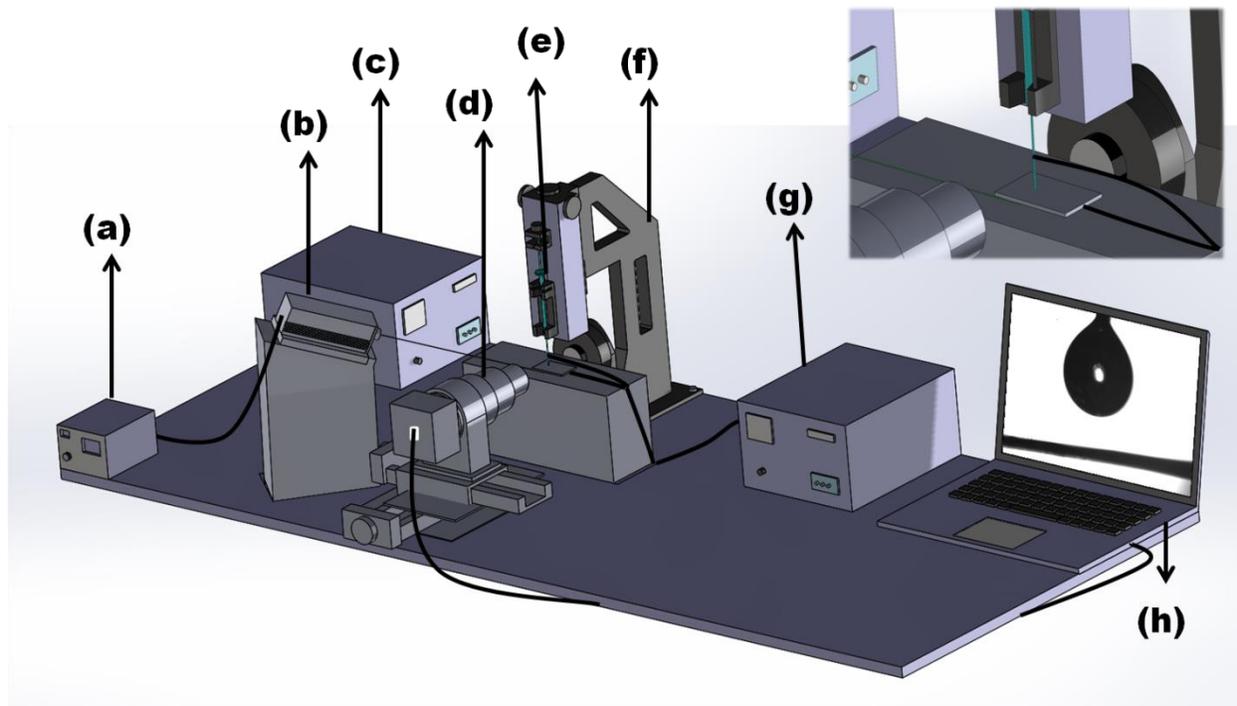

**FIG. 1:** Schematic of the experimental set up, comprising of (a) laser controller (b) laser mounted on stand, with cylindrical lens for light sheet (not illustrated) (c) droplet dispenser and backlight controller module (d) CCD camera with two-axis movable stand and mounted with microscopic lens assembly (e) sterile glass syringe (f) droplet dispensing mechanism (g) programmable AC power source (h) data acquisition computer. The inset illustrates a zoomed in view of the arrangement of electrodes with respect to the needle. The whole setup is housed within an acrylic chamber and mounted on a vibration-free table top.

The evaporation process is recorded via a moderate frame rate CCD camera, mounted with a microscopic lens (Holmarc Opto-mechatronics, India). The recording frame rate is kept at 10 fps, with a resolution of 1280 x 960 pixels. A brightness controlled LED backlighting arrangement (DpLED, China) is employed as the light source. The ambient temperature and humidity are recorded using a digital thermometer and a digital hygrometer. The temperature and humidity range from 25 ± 3 $^{o}$C and 54 ± 4 %, respectively, at a location of 5 cm away from the droplet. The whole setup is housed within an acrylic chamber to isolate it from atmospheric and human induced aberrations. The video recorded is sliced into an image stack using the open source image processing software ImageJ. A macro subroutine was written to extract physical data from the image stack. Each instantaneous droplet was fit to the closest pendant drop shape, and the equivalent diameter was obtained from the drop shape analysis. To quantify and visualize the flow patterns within the droplet, if any, Particle Image Velocimetry was employed. The salt solution was seeded with neutrally buoyant fluorescent particles (polystyrene, ~ 10 μm, Cospheric LLC, USA). For the PIV studies, the backlight is dimmed for a 90 seconds interval and the laser (Roithner Gmbh, Germany) is used as the illumination source. A plano-convex lens is used to generate a laser sheet (~ 0.5 mm thickness) to illuminate mid-plane of the droplet. The laser emits at 532 nm wavelength and is used at 10 mW peak power condition. The PIV is performed during the very initial stages of evaporation (within first 5 minutes) so that internal circulation effects due to evaporation induced change in the concentration are not captured. . The frame rate for the PIV is maintained at 30 fps and a minimum resolution of ~ 120 pixels/mm is ensured. Post-processing is performed using the open source code PIVLab.. A cross-correlation algorithm with 3 pass interrogation windows of 64 pixels, 32 pixels, and 16 pixels has been used. A stack of 600 images are investigated for maximizing the signal-to-noise ratio in the data and spatially averaged velocity contours and fields are obtained.

## 3. Results and discussions

### 3.1. Evaporation kinetics in the presence of electric field

Initially, the evaporation kinetics of pure water and variant salt solutions are noted in the absence of electric field, and these cases behave as the controls. Subsequently, evaporation of water under variant electric field strengths are experimented upon, and no deviations from the zero-field case is observed, which establishes that the field has no governing effect on the evaporation kinetics of only the pure fluid. Further, the evaporation characteristics of the saline droplets under variant field strength and frequency are studied. Figure 2 (a) illustrates the evaporation behavior of pure water and 0.01 M NaI solution under the effect of electric field of 50 Hz and different strengths as a function of evolving time. In a manner similar to the pure water or saline solution in absence of electric field, the evaporation kinetics of the saline droplets under the influence of electric field is observed to conform to the classical $D^2$ law, expressible as

$$\frac{D^2}{D_0^2} = 1 - k \frac{t}{D_0^2} \tag{1}$$

Where, D, $D_0$, t, and k represent the instantaneous and initial diameter of droplet, the elapsed time and the evaporation rate constant, respectively. The best-fit straight line is implemented to the evaporation curves (Fig. 2 (a)) and the slope of the curve yields the evaporation rate constant, in accordance to Eqn. 1. It is evident from Fig. 2 (a) that the addition of salt enhances the evaporation rate of the water. The enhancement of evaporation due to addition of salt is caused by solute-thermal advection within the droplet and has been reported in literature [14]. With the field switched on, the evaporation rate is observed to decrease drastically, and deteriorates even below the rate of evaporation of the pure water. However, at higher field strengths, the evaporation enhances by a small degree, but is still below the rate of water, or equivalent. Figure 2 (b) illustrates evaporation rate for the cases corresponding to Fig. 2 (a). The water droplet shows no change in evaporation rate under field effect (not shown in figure).

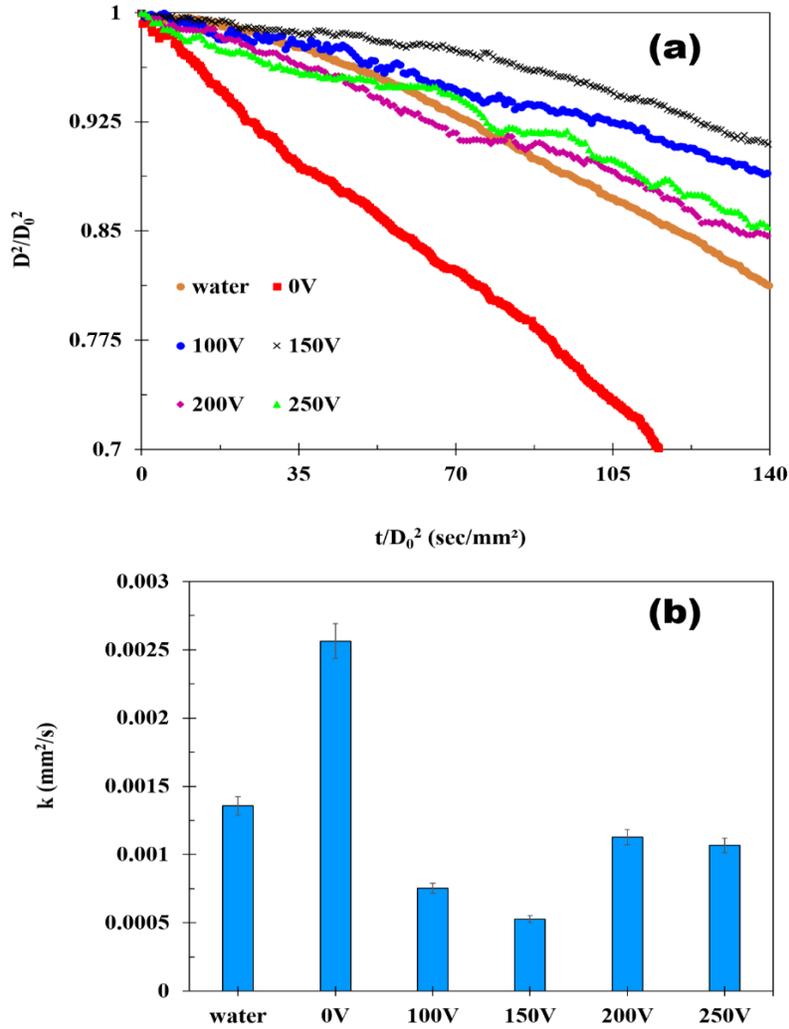

**FIG. 2.** (a) The evaporation dynamics of NaI solution (0.01 M) at 50 Hz and different field strengths (b) Evaporation rate constant corresponding to field strength variation.

Figure 3 (a) compares the evaporation rates for two salt (NaI and $CuSO_4.5H_2O$) solutions at the same concentration (0.1 M), under the influence of two different field frequencies (50 and 200 Hz) at a field strength of 150 V. For comparison, the case of pure water has also been illustrated. The figure caters to two comparisons, one for change in frequency and the other for nature of salt solution and figure 3 (b) illustrates the evaporation rates for cases corresponding to 3 (a). It is observable that the evaporation of $CuSO_4$ is faster compared to NaI, and this is already explained in the previous report by the authors [14, 23]. The hydration molecules of $CuSO_4$ leads to such erratic behavior [14], however, NaI exhibits reliable observations. It is observed from Fig. 3 (b) that increase in the field frequency leads to improved evaporation rates, whereas increase in field strength leads to deteriorated evaporation (fig. 2). This observation is further clarified later from fig. 5. The decrease in evaporation rate in the presence of electric field maybe explained based on the change in the ionic orientation within the polar fluid with increase in field strength. In a strongly polar fluid such as water, the solvated ions form physical bonds with the polar water molecules, with the cations partially attracted to the partly polar hydrogen of water, and the anions to the oxygen. The solvated anions and cations aligns themselves along the field lines at the onset of the electric field, and the electrostatic force on the solvated ions increase with increment in potential difference across the electrodes. This additional body force on the solvated ions strengthens the physical bond with the polar water molecules [24], thereby increasing the threshold energy required by the water molecules at the droplet-air interface to diffuse away from the droplet, thereby decelerating evaporation rate. The frequency of the field on the other hand determines ionic migration scope of the ions. At higher frequencies, the field polarity reverses at faster rates, which causes the solvated ions to fluctuate about their mean positions at higher frequencies, in addition to the thermal fluctuations within the system. This increased field induced fluctuation leads to weakening of the physical bonds with the polar water molecules, which eases the probability of the molecules at the interface to diffuse into the air phase, thereby improving the evaporation rate.

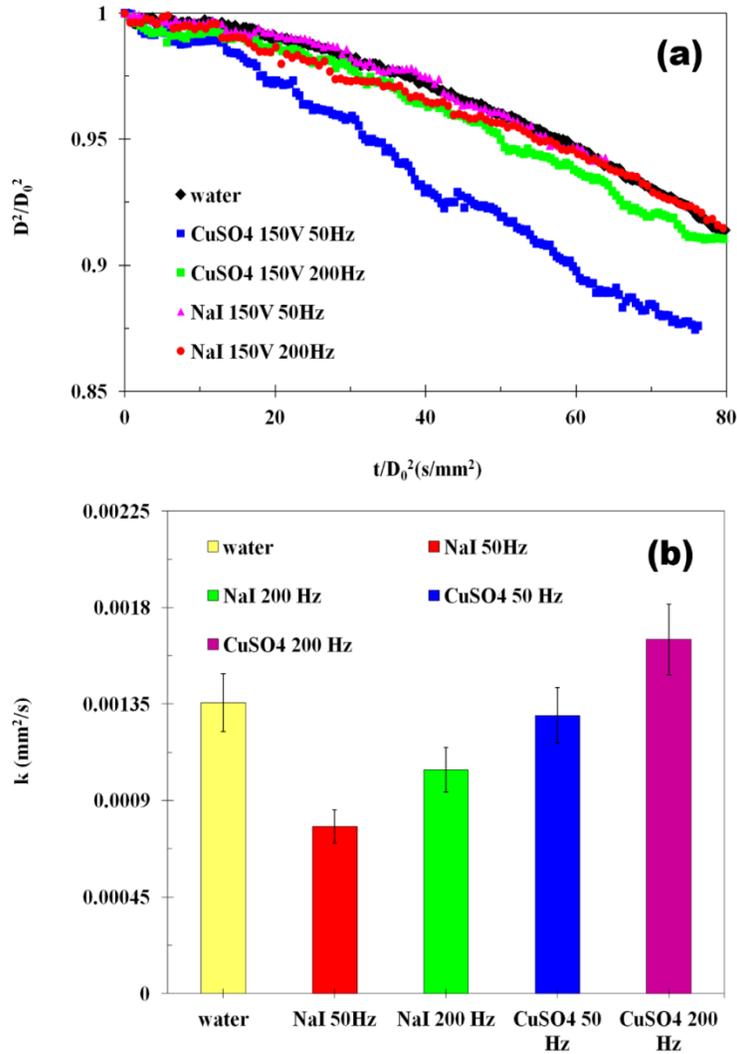

**FIG. 3:** (a) Evaporation dynamics of different salt solutions (0.01 M) at 150 V and for different field frequencies (b) Evaporation rate constants corresponding to frequency variation.

### 3.2. Role of surface tension on evaporation in electric field ambience

While a qualitative reasoning for arrest of evaporation rate has been provided, the process needs to be discussed based on the fluid dynamics, thermal and species transport aspects. The surface tension of the fluid with respect to its ambient phase determines the evaporative rate from the droplet and its behavior in a field environment may be important to explain the field influenced evaporation dynamics. The surface tension of the droplet is measured using the pendant drop method under variant field conditions.  Figure 4 illustrates  the response of surface tension to stimuli of changing field strength (at constant frequency) and changing frequency (at constant field strength). The surface tension of the pure water is noted as 72.5 mN/m and that of the 0.2 M NaI solution is observed to enhance to 82 mN/m, which is consistent with reports on salt solutions [14]. . With increase in the field strength, the variation in surface tension ranges from 82 to  77 mN/m, whereas, in case of increase in field frequency,  the variation ranges from 82  to 62 mN/m. Additionally, the decrease  in case of increase in field strength is gradual whereas for increase in frequency,  the behaviour is steeper However, the matter of importance is that increase in both field strength and frequency leads to decrease in the surface tension value of the saline solution, which signifies that evaporation rates should augment due to reduction in threshold surface energy for the departing water molecules. While this partially explains the improvement in evaporation rates with increase in frequency, it falls short to explain the deteriorated evaporation rates upon increase of field strength. Thereby, additional physical probing is essential to establish the dominant mechanism at play.

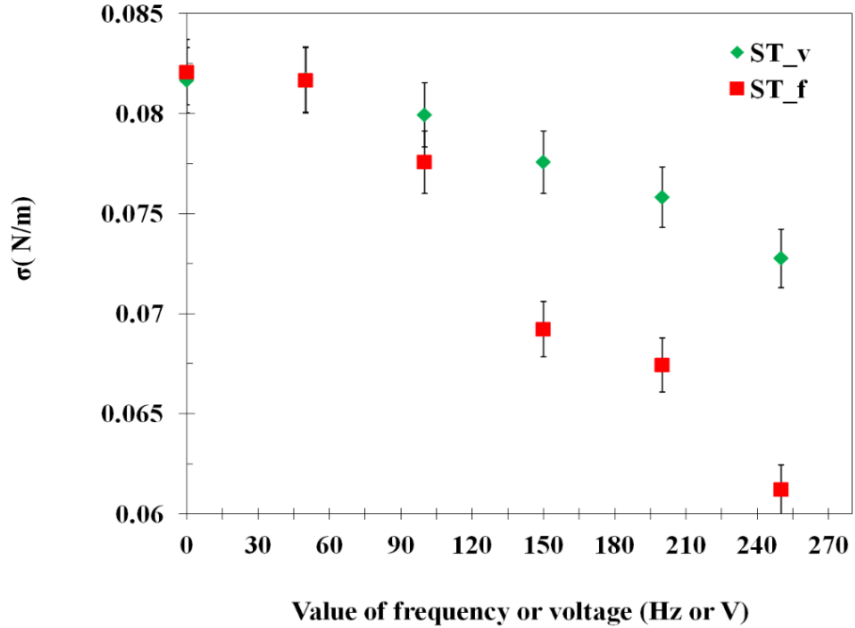

**FIG 4:** The variation of surface tension with change in voltage (indicated as v) and frequency (indicated as f) for 0.2 M NaI solution. The extreme limits of change in the surface tension have been illustrated

### 3.3 Contribution of diffusion dominated evaporation

As multiple mechanisms could be involved in modulating the evaporation of saline pendant droplets under the influence of electric field, a sequential discussion on each dominant mechanism provides clarity regarding their contributions. The basic mechanism by which a pendant droplet evaporates into the ambient phase is by molecular diffusion. Consequently, the diffusion driven evaporation model [25] is appealed to. The approach considers a thin film of vapor formed on the surface of the droplet by the escaping water molecules, from which the vapor further escapes to the ambient phase by virtue of species concentration gradient. The approach considers evaporation to be quasi-steady; and the thermal and mass transport is through the diffusion film. Applying conservation equation on uniformity of droplet temperature and the temperature and vapor concentration of the diffusion layer outside droplet as boundary conditions, the evaporation rate is expressed by the eqns. 2-5 as

$$B_M = \frac{Y_s - Y_\infty}{1 - Y_s} \tag{2}$$

$$B_T = (1 + B_M)^\phi - 1 \tag{3}$$

$$\phi = \frac{C_{pf} Sh}{C_{pg} Nu Le} \tag{4}$$

$$\dot{m} = 2\pi \rho_g D_v R \ln(1 + B_M) Sh = \frac{2\pi \lambda R}{C_{pf}} \ln(1 + B_T) Nu \tag{5}$$

In eqns. 2-5, $\dot{m}$ is the droplet evaporation rate, $D_v$ is the diffusion coefficient of the vapor with respect to the ambient phase, $\rho_g$ is the density of the ambient gas, R is the droplet radius, $\lambda$ is thermal conductivity of the ambient gas phase, $C_{pf}$ and $C_{pg}$ are the specific heat of the vapor film and surrounding gas. $Nu$, $Le$, $Sh$, $B_M$ and $B_T$ are the Nusselt number, Lewis number, Sherwood number, Spalding mass and heat transfer numbers, respectively. $Y_s$ and $Y_\infty$ are the mass fractions of the vapor at the droplet surface and in the ambient phase. Figure 5 illustrates the comparison of the diffusion driven model predictions with experimentally obtained evaporation rate constants. It is observed that the diffusion driven model is incapable to predict the evaporation rate values under different field constraints. However, it may be noted that the model is able to predict the evaporation rate for pure water case (not illustrated). This observation is justified as the diffusion driven model only considered phenomena across the vapor diffusion layer and does not encompass mechanisms at play within the droplet or at the interface. Thereby, it is evident that internal dynamics of the droplet as well as interfacial dynamics need to be accounted for explaining the arrested evaporation kinetics.

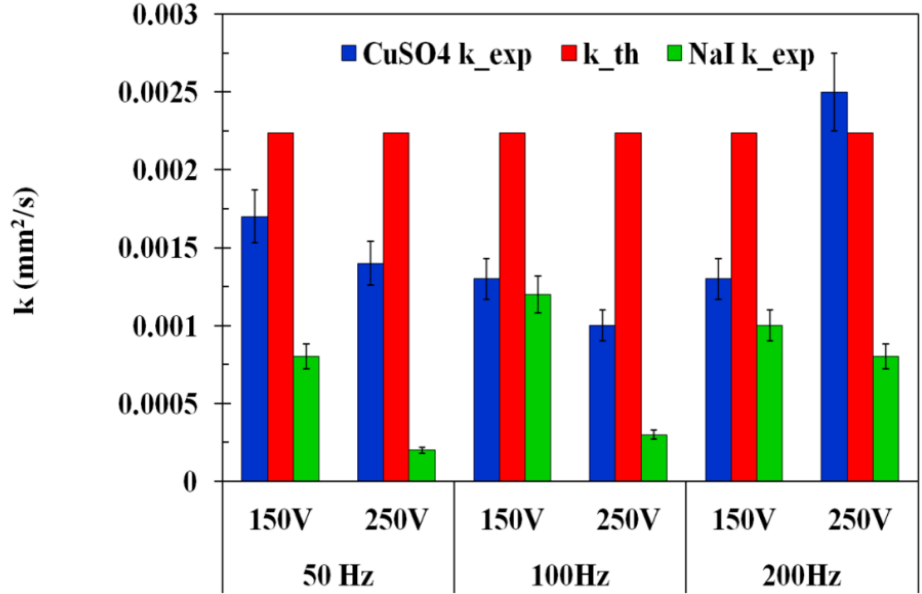

**FIG. 5:** Comparison of experimental evaporation rate with the diffusion driven theoretical approach [25]. The diffusion driven model fails to predict the experimental evaporation dynamics.

### 3.4. Internal advection dynamics under electric field stimulus

Internal circulation or advection within droplets, caused by thermal and/or solutal gradients, has been reported to be a major phenomena towards change in evaporation rates. The advection within the droplet and along the droplet interface, caused by thermo-solutal gradients, generates shear within the shrouding diffusion layer. This shear leads to replenishment of the saturated diffusion layer with ambient phase gas, which enhances the species transport caliber, thereby improving evaporation [14, 22, 26]. The internal circulation pattern is qualitatively and quantitatively determined PIV experiments and the effects of field frequency and strength are deduced. PIV studies were done within the initial 5 minutes of initation of evaporation to avoid the effects of change in the concentration of the solution on the internal advection kinetics.

Visulaization studies have been performed for 0.2 M NaI solution and water droplets. The control cases of water with and without field stimulus and the salt solution droplet with no electric field are obtained to segregate the effects of solutal advection [14, 26], field frequency and strength. The internal circulation dynamics for water and salt solution droplet in zero-field environment conforms with reports in literature [14, 22]. The water droplet shows no appreciable change in advection dynamics in field environment. Figure 6 illustrates the time-averaged velocity contours and fields (for 600 subsequent velocity fields) for different field strength and frequency configuration. Fig. 6 (a) shows the contours for 0.2 M NaI under zero-field conditions and exhibits a major advection cell within the entirety of the bulk of the droplet. When the field frequency is invariant and the field strength increased (fig. 6 (b) and (c)), the vortical motion changes orientation to an orthogonal plane, and two opposing vortices form near the bulb of the pendant at 100 V, alongwith a supression in circulation velocity. At 250 V, the circulation is further supressed, and a single small circulation cell operates at the bulb of the pendant. Infact, under high field strength case of 250 V, the spatially averaged velocity is almost equivalent to the minute drift noted within water droplets [14, 26], with the exception of the region of circulation near the bulb.

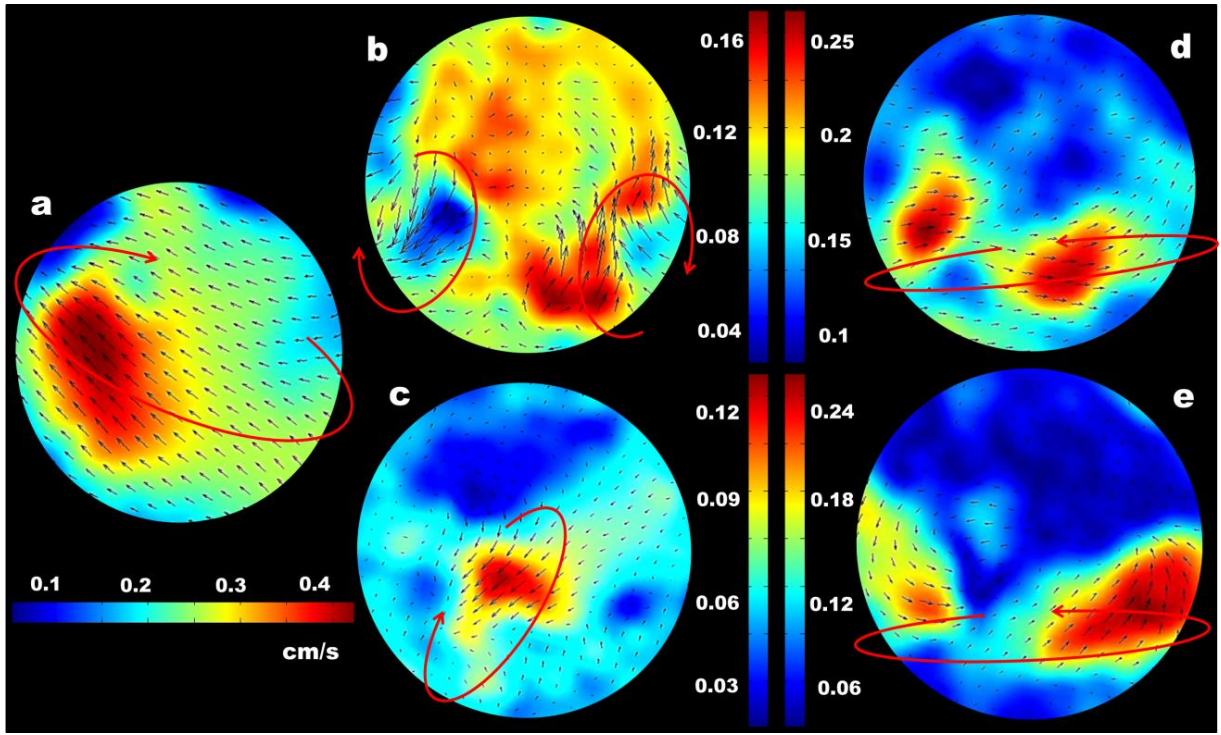

**FIG. 6.** Internal circulation velocity contours and time-averaged flow field for (a) 0V (b) 50 Hz, 100 V (c) 50 Hz, 250V (d) 200 Hz, 100 V (c) 200 Hz, 250V respectively. The concentration of NaI is 0.2 M for all cases. The large red arrows indicate the spatially dominant direction of advection.

In the event of increase in frequency at constant field strength (figs. 6 (b) and (d)), it is observed that the pattern of advection regains similarities with the zero-field case, albeit with the circulation region smaller in size than the zero-field case. Additonally, the velocity magnitude, although less than the zero-field salt solution case, is relatively greater than the high field strength case. A quick introspection of the cases in figs. 6 (a) – (d) reveals that the field strength has a dominating role towards arrest of internal circulation in comparison to field frequency. However, unlike the case of high field strength, high frequency cases lead to reduction in the internal circulation velocity, but without major changes in the advection patterns. As observable from fig. 6 (d) and (e), at high frequencies, the internal circulation is very similar to the zero-field case, however, the major portion of advection is again towards the bulb of the droplet. This behaviour is in all probabilities due to the nature of the field imposed. While the bottom electrode is plate type, the top electrode is the needle itself. Thereby the field liens diverge from the tip of the needle and spread onto the bottom electrode. Thereby, the field lines influence the neck of the pendant to lesser extent than the bulb, which is possibly the reason why internal advection dynamics are concentrated towards the bulb in all field conditions. It has been reported that internal advection (thermal, solutal, or otherwise) causes interfacial shear, which replenishes the vapour diffusion layer with ambient phase, leading to enhanced evaporation in saline or binary mixtures [22]. Visualization reveals that the electric field arrests the advection, which in turn diminishes the evaporation rate of the saline solution to that of the water case. Moreover, as discussed previously, the enhanced electric body force on the water molecules due to physical bonds with the solvated ions further poses additional energy barrier at the interface for the escaping water molecules, which further deteriorates the evaporation rate in certain cases (fig. 2).

Fig. 7 (a) illustrates the temporal spectrum of the spatially averaged instantaneous circulation velocities under different field constraints for 30 sec duration. The water case shows minor drift, whereas the zero-field salt case exhibits largely augmented circulation, with large-scale temporal fluctuations in the average velocity, and this corroborates with reports for saline or binary fluid systems [14, 22, 26]. In one case, field at 50 Hz and 100 V has been applied for the first 15 sec, and thereafter, the field frequency is increased to 200 Hz at same field strength. In another case, initially 50 Hz and 100 V has been applied, and after 15 sec the field strength has been increased to 250 V at same frequency. It is observed that initially, both cases exhibit similar spectra, since the field strength and frequency are same. When the field is increased suddenly at 15 s from 100 to 250 V, a reduction in the spatially averaged velocity is noted, with the average velocity dropping by ~70 % of the original value. However, in the event the frequency is increased from 50 to 200 Hz, the reduction in the circulation velocity is only 25 %. Additionally, it is observed that the higher field strength case also dampens out the temporal fluctuations largely compared to the higher frequency case. At higher frequencies, the electrophoretic motion of the solvated ions exhibit field influenced fluctuations, which lead to local hydrodynamic fluctuations, thereby leading to augmented temporal velocity fluctuations. The higher field strength however is able to arrest such local fluctuations due to enhanced electric body force, leading to smoothened temporal fluctuations in velocity. The time spectra and field modulation clearly illustrates the dominant role of the field strength over the field frequency, as discussed previously.

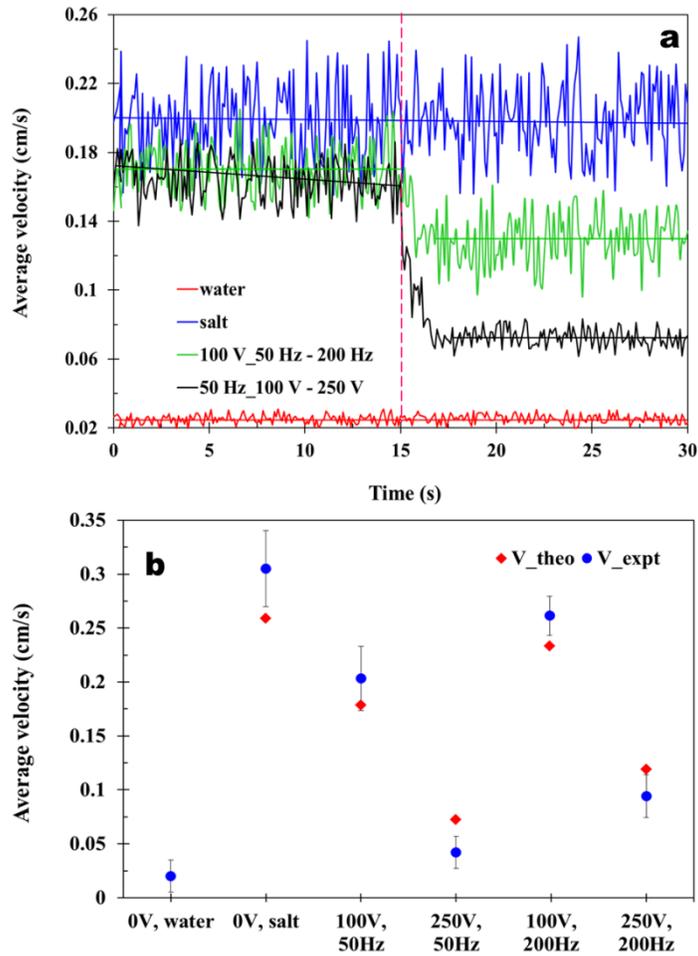

**FIG. 7.** (a) Spatially average circulation velocity time series spectra for water and 0.2 M NaI salt solution droplets under variant field parameters. The straight lines illustrate the best-fit average velocity (b) Comparison of the theoretical velocities from the scaling model with respect to the spatially averaged, experimentally determined velocities.

### 3.5. Scaling analysis of electro-thermal advection

The flow visualization studies reveal that modulation of the internal advection dynamics by the field features leads to arrest of the evaporation rates in the saline pendant droplets. Inter advection in such droplets may be caused by thermal advection or solutal advection, or a combination of both. Consequently, it becomes pertinent to deduce which modality of advection is affected by the electric field and the qualitative extent by which it is affected. A scaling analysis has been put forward to segregate the dominant mechanism from the recessive and to quantify the characteristics of the same, along lines of the philosophy appealed to in literature [14, 26]. . The scaling analysis is based upon assumptions that thermophysical properties remain invariant with respect to the minor changes in humidity and temperature within the experimental system , and surface tension only varies due to either thermal or concentration gradients generated within the droplet. The electric field is assumed to have no effect on the thermophysical properties of the fluid (since the field strengths employed are moderate and not in kV range). The energy balance equation applied to the droplet during evaporation and under the influence of thermally driven advection within and influence of electric field is expressible as

$$\dot{m} h_{fg} = k_{th} A \frac{\Delta T_m}{R} + \rho C_p U_{c,m} A \Delta T_m - \rho C_p V_f A \Delta T_m \qquad (6)$$

The variables $\dot{m}$, $h_{fg}$, $k_{th}$, $A$, $\rho$, $C_p$, $\Delta T_m$, $U_{c,m}$ and $V_f$ represent the mass rate of evaporation, the latent heat of vaporization, the thermal conductivity of the fluid, the heat transfer area of the droplet, density and specific heat of the fluid, temperature difference across the interface of the droplet and the droplet bulk, the internal circulation velocity due to thermally driven advection and internal velocity with electric field driven advection, respectively. In eqn. 6, the left hand side represents energy flux associated to the evaporative flux from the droplet. The right hand side terms represent of diffusion heat transfer across the droplet, the internal advective term due to temperature gradient driven surface forces, and the electro-thermal advection term, respectively. The thermal advection arises due to the difference of temperature across the droplet bulk and interface, which is brought about by evaporative cooling [22]. The additional electro-thermal component represents the advection due to modulation of the circulation dynamics in presence of electric field. The negative sign represents the experimental observation of arrested evaporation rates in presence of field. The thermal gradients generated by the evaporation drives the thermal advection, both within the bulk and at the interface, caused by the thermal Marangoni effect. However, since the interfacial dynamics cannot be visualized and quantified directly, the effective circulation dynamics at the interface and the bulk can be quantified using the Marangoni number. The spatially averaged internal circulation velocity due to thermal Marangoni effect is expressed as $U_{c,m} = \frac{\sigma_T \Delta T_M}{\mu}$ as [22], where $\sigma_T$ represents the rate of change of surface tension due to change of temperature and $\mu$ is the kinematic viscosity of the fluid. The average electro-thermal advection velocity $V_f$ deduced from force balance of the electrohydrodynamic system within the droplet environment. The solvated ions within the fluid experience motion due to the thermal advection, which in the presence of an electric field gives rise to a modified velocity. For a moving charged entity in presence of the electromagnetic field, the net force experienced is expressible as

$$\vec{F} = q\vec{E} + (\sigma_e \cdot \vec{E} \times \vec{B}) + \sigma_e (\vec{v} \times \vec{B}) \times \vec{B} \tag{7}$$

where $\vec{F}, q, \vec{E}, \vec{B}, \vec{v}$, and $\sigma_e$ represent the electromagnetic force, the charge of the ion, the electric field strength, the magnetic field strength, the velocity of the particle and electrical conductivity of the fluid, respectively. In the case of only electric field and no magnetic field, eqn. 7 reduces to

$$\vec{F} = q\vec{E} \tag{}$$

Considering the volumetric body force, eqn. 8 can expressed as

$$f = \rho_e \vec{E} \qquad ()$$

where $\rho_e$ is the charge per unit volume ($=zeN$) and $f$ is the electromagnetic force per unit volume, and $z$, $e$ and $N$ represent valence of the solvated, the elementary charge magnitude and the number density of solvated ions in the system. The electrical body force leads to inertial advection within the droplet, which can be expressed as

$$\rho_e E = \rho a \qquad ()$$

where $a$ is acceleration component due to in circulation velocity due to applied external electric field. The acceleration can be scaled as $a \sim V_f/t$ and further the time elapsed can be scaled as $t \sim R/V_f$, where $R$ is the instantaneous radius of the droplet. Therefore, the eqn. 10 can be expressed as

$$\rho_e E \approx \rho \frac{V_f^2}{R} \qquad ()$$

From eqn. 11, the internal circulation velocity in the presence of electric field stimulus can be expressed as

$$V_f \approx \sqrt{\frac{\rho_e E R}{\rho}} \qquad ()$$

The product of electric field and the associated length scale (instantaneous droplet radius) is expressed as the electric field potential (V) and eqn. 12 can be expressed as

$$V_f \approx \sqrt{\frac{eNzV}{\rho}} \approx \frac{v}{R}\sqrt{\frac{eNzVR^2}{\rho v^2}} \qquad ()$$

$$V_f \approx \frac{v}{R}\sqrt{E_{HD}} \qquad ()$$

where $v$ and $E_{HD}$ represent the kinematic viscosity and the Electrohydrodynamic number, respectively. Substituting the respective expressions for $U_{c,m}$ and $V_f$ in eqn. 6, and arranging, the eqn. 17 is obtained

$$\rho \dot{R} A h_{fg} = k_{th} A \frac{\Delta T_m}{R} + \rho C_p A \Delta T_m \frac{\sigma_T \Delta T_m}{\mu} - \rho C_{pf} A \Delta T_m \frac{v}{R}\sqrt{E_{HD}} \qquad ()$$

$$\rho \dot{R} R h_{fg} = k_{th} \Delta T_m [1 + \rho C_p \frac{R\sigma_T \Delta T_m}{k_{th}\mu} - \rho C_{pf} \frac{v}{k_{th}}\sqrt{E_{HD}}] \qquad ()$$

$$\rho \dot{R} R h_{fg} = k_{th} \Delta T_m [1 + Ma_T - \Pr \sqrt{E_{HD}}] \qquad ()$$

where $Pr$ and $Ma_T$ represent the Prandtl number and thermal Marangoni number, respectively. For stable internal advection, $Ma_T \gg 1$ [22], and eqn. 17 can be scaled as

$$\rho \dot{R} R h_{fg} \approx k_{th} \Delta T_m [Ma_T - \Pr \sqrt{E_{hd}}] \qquad ()$$

In eqn. 18, the second term on the right hand side represents the effect of the electric field on the thermal diffusion within the droplet and is termed as the electro-Prandtl number, and as observed front the eqn. 18, acts as a deterrent to the thermal Marangoni advection dynamics (represented by the thermal Ma). The whole term $(Ma_T - \Pr\sqrt{E_{hd}})$ is termed as the effective electro-thermal

Marangoni number ($Ma_{t,e}$), and its modulus value determines the strength of electro-advection within the droplet. The extent of deterioration in internal velocity depends on the magnitude of the electro-Prandtl number, which is a direct function of the field parameters.

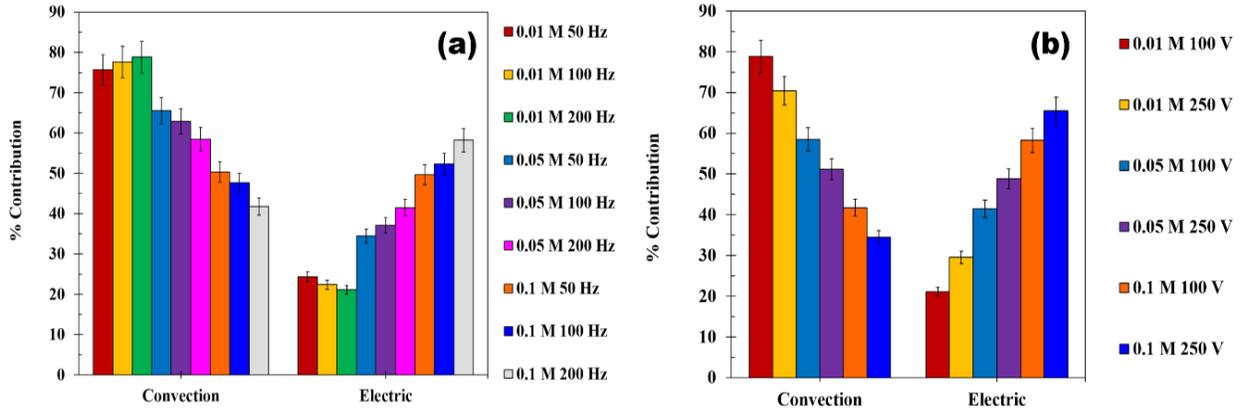

**FIG. 8:** The contribution of different mechanisms in thermal transport with (a) frequency variation at 150V and (b) voltage variation at 100 Hz

The possibility that thermal gradients within the droplet lead to circulation due to Rayleigh convection also exists and its potential vis-à-vis the Marangoni convection must also be established. The internal circulation velocity, if induced by the buoyancy effect, can be expressed as $u = g\beta\Delta T_R R^2/\nu$ [22], where $g$, $\beta$ and $\Delta T_R$ are acceleration due to gravity, the thermal expansion coefficient of the fluid and the thermal difference within the droplet due to which buoyancy advection effect might arise. The temperature difference due to the Rayleigh convection can be thereby expressed as [14, 22]

$$\Delta T_R = \sqrt{\frac{\nu \dot{R} h_{fg}}{g \beta R^2 C_p}} \qquad ()$$

where $\dot{R}$ is the rate of change of droplet radius. The associated Rayleigh number is expressed as

$$Ra = \frac{R^2}{\alpha} \sqrt{\frac{\dot{R} h_{fg} g \beta}{C_p \nu}} \qquad ()$$

However, the contribution of buoyancy driven circulation in evaporation process is negligible as compared to Marangoni convection, and this has been confirmed in literature [22, 26]. Additionally, it is of importance to determine if the circulation patterns within the droplet are temporally stable or are in general intermittent. This can be determined from the stability criteria (Nield [27] and Davis [28]) represented in the form of a summation of two ratios, $Ra/Ra_c$ and $Ma/Ma_c$. The $Ra_c$ and $Ma_c$ are the critical Ra and critical thermal Ma. The ratio $Ra/Ra_c$ presents the extent of thermal Rayleigh advection relative to its possible maxima within the droplet and $Ma/Ma_c$ depicts the same for the thermal Marangoni advection. The criteria for temporally stable circulation is

$$\frac{Ra}{Ra_c} + \frac{Ma}{Ma_c} = 1 \qquad ()$$

The $Ma_T$ and Ra values have been plotted for variations in frequency and field strength in figure 9 (a) and (b), respectively. The addition of salt results in shifting of the points towards the right and upwards on the $Ma_T$ vs Ra plot compared to pure water case. The zero field cases show marginally stable circulation (lies between the two stability criteria lines), which enhances evaporation compared to water. With the input stimulus of different frequencies (fig. 9 (a)) at 150 V, the points are observed to ubiquitously shift downwards and below the region of stable circulation. This justifies the decreased velocity of circulation observed in the case of field influenced evaporating droplets during flow visualization. Several cases also drop below the water case, which signifies that the thermal Ma effect is arrested largely by the electric field and as observed before, higher frequiencies lead to lower degrees of arrest compared to lower frequencies. For variation of field strength (fig. 9 (b)) at 100 Hz, all points lie within the criteria of unstable circulation, and in comparison to fig. 9 (a), the thermal Ma values are further lowered. This reveals that the field strength has a domineering role over frequency in arresting internal circulation due to thermal advection, and complies with the visualization results.

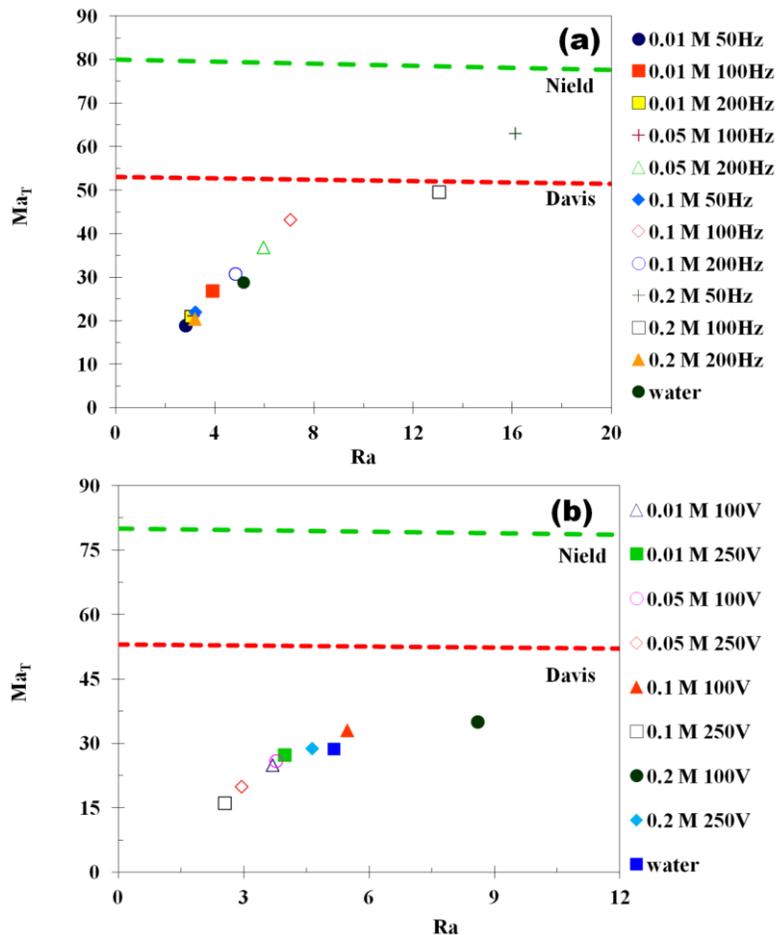

**FIG. 9:** Stability plot for thermal Ma with respect to Ra to determine nature of temporal stability of internal circulation (a) with variation of frequency and salt concentration(b) with variation of field strength and salt concentration.

A similar stability plot representing the effective electro-thermal Ma ($Ma_{t,e}$) and Ra has been illustrated in figure 10. Figure 10 (a) illustrates the variation of frequency And has been segregated into two regimes. The green region represents the condition where the thermal Ma iis greater than the electro-Prandtl number $\left(Ma_T > Pr\sqrt{E_{HD}}\right)$, whereas the pink region represents $\left(Ma_T < Pr\sqrt{E_{HD}}\right)$. For stable circulation to exist, the points should cross the stability lines (not shown in pink regions) in the corresponding region. Iso-$E_{HD}$ lines have been illustrated in figure 10, which show the variation and trend of internal circulation kinetics for known values of $E_{HD}$. It is observed that all the points in figs. 10 (a) and (b) are further shifted downwards and leftwards on the map (in comparison to their locations on fig. 9), which signifies that the effective electro-thermal Ma is the main modulating factor for the thermal advection within the droplet. It is noteworthy that in several cases, such as $CuSO_4$ 0.05 M at 100 V and NaI 0.05 M at 250 V exhibit very similar circulation velocities in the visualization exercise, but their positions on fig. 10 (a) with respect to the $Ma_{t,e}$ are quite different. While the copper sulphate case lies in the green region, the sodium iodide case lies in the pink zone. This provides and interesting insight on the electrohydrodynamics of the problem at hand. While both droplets exhibit similar internal advection velocity and similar evaporation dynamics (of course NaI has to be at higher field strength since its zero-field advection is stronger [14]), the copper sulphate case undergoes internal advection by virtue of superior thermal Marangoni advection, whereas the sodium iodide case experiences circulation due to superior electro-Prandtl component of advection. This essentially signifies the arrest of evaporation is due to competitive electro-thermal and thermal

Marangoni advection. In the event the electro-thermal component largely exceeds the thermal Ma component, it can lead to large value of $Ma_{t,e}$ in the pink region, which will give rise to augmented circulation, whose stability will be determined from the criteria by Davis and Nield in the pink region (not illustrated). The iso-$E_{HD}$ lines provide further insight to the problem. Assuming a saline droplet exhibits circulation and lies somewhere near the Davis line (fig. 10) at zero-field. At a constant Ra, if the $E_{HD}$ is enhanced for a saline solution (by improving field strength), the value of $Ma_{t,e}$ will deteriorate to zero, with arrest in circulation and the evaporation rate. With further increasing field strength, the electro-Prandtl number overshadows the thermal Ma, causing the absolute value of the $Ma_{t,e}$ to increase once more, leading to increased circulation (due to electro-advection) and increased evaporation rate. Similar instances are observable from fig. 10 (a) and (b).

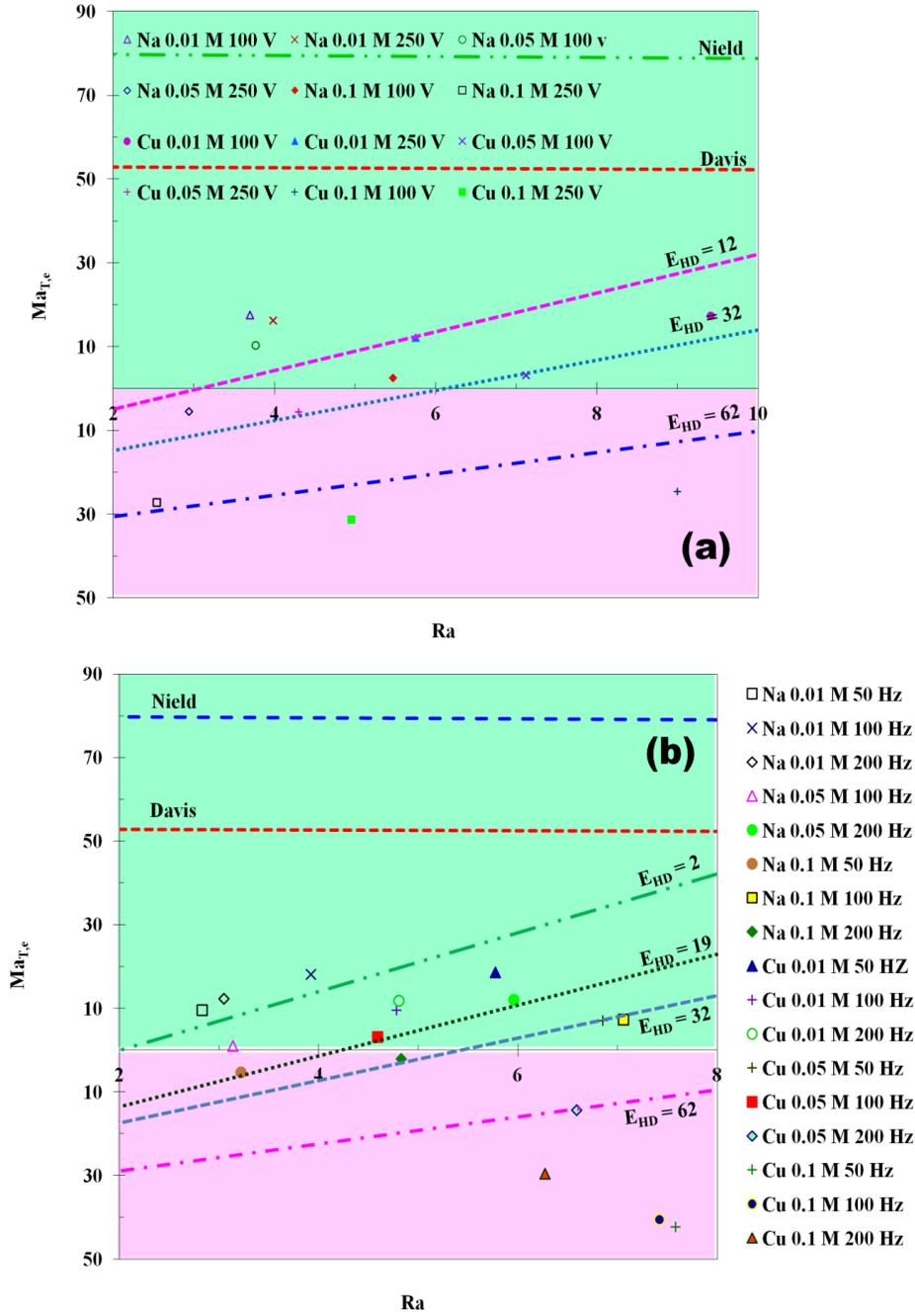

**FIG. 10**: Stability plot for electro-thermal Ma with respect to Ra (a) with variation of field strength and concentration (b) with frequency and concentration variation. The iso-$E_{HD}$ lines represent the shift of the points of stability under variant field conditions.

## 3.6. Scaling analysis of electro-solutal advection

Although a detailed analysis and discussion on the role of electro-thermal advection has been presented, literature shows that saline droplets exhibit enhanced evaporation due to the domainant solutal advection over the thermal one [14], thereby the possible role of the electro-solutal advection must also be analysed. As evaporation of droplet with solvated ions progresses, only fluid molecules depart with time, thereby enhancing the bulk concentration of solvated ions as the process evolves. However, it is noteworthy that the fluid molecules depart only from the interface, which leads to localized enhanced solute concentration at the droplet interface in comparison to the bulk of the droplet. This concentration gradient leads to a solutal component advection within the droplet. The interface also experiences solutal Marangoni advection caused by the concentration gradient along the droplet interface (due to the inherent difference in evaporation created by the very geometry of the pendant shape) [14, 26]. The quantification of evolving bulk concentration is achieved from total species conservation (VC=constant, where V and C represent the instantaneous volume and dynamic bulk concentration). For mapping the dynamic interfacial concentration, initially the relationship between the surface tension with respect to the concentration is determined [26]. Further, the instantaneous surface tension of the droplet is mapped as the evaporation process evolves, via drop shape analysis. Correlating the instantaneous surface tension with the concentration relation yields the dynamic evolution of the interfacial concentration. The exercise is done only for the initial 15-20 mins of the evaporation process. Beyond this the Worthington number (Wo) of the droplet ($Wo=V_{inst}/V_0$, where V is the droplet volume, and subscripts *inst* and *0* represent instantaneous and initial conditions$_)$ reduces below ~ 0.75, and further determination of surface tension by drop shape analysis will be erroneous. The analysis provides distinct roots as the interfacial concentration, and the real root is considered. Figure 11 illustrates the behaviour of bulk and interfacial concentrations for NaI solution droplet (0.2 M) under different field constraints.

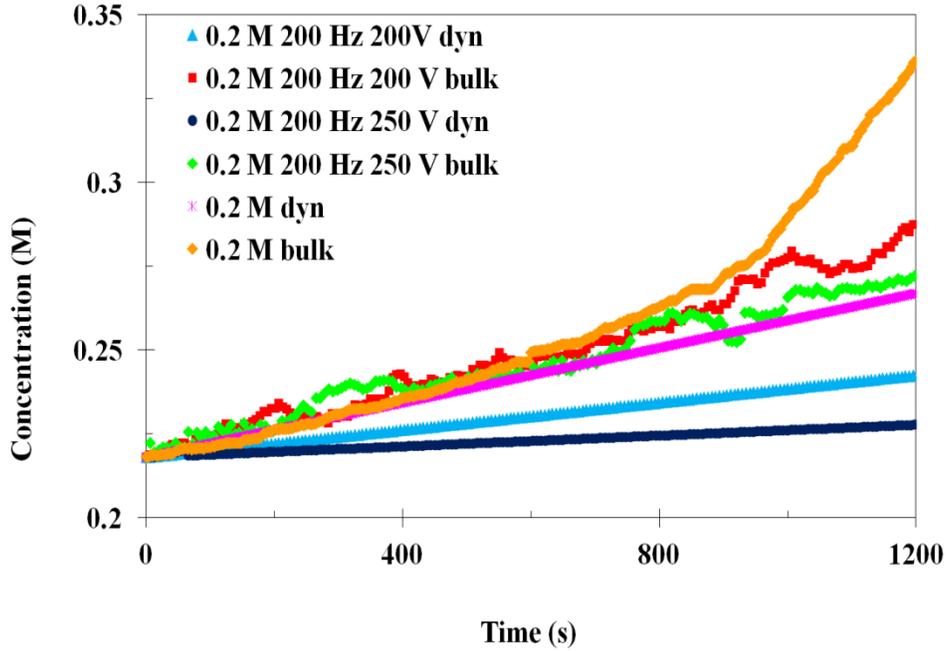

**FIG. 11:** The dynamic and bulk concentration within the droplet (0.2 M NaI) as evaporation evolves. The '*int*' represents the dynamic surface or interfacial concentration and '*bulk* ' represents the dynamic bulk concentration.

To quantify the effect of electro-solutal advection on the evaporation kinetics, species conservation is applied as

$$\dot{m} = DA\frac{\Delta C_m}{R} + U_{c,m}\Delta C_m A - V_{f,c}\Delta C_m A \qquad ()$$

where $\dot{m}$, $D$ and $\Delta C_m$ are the mass rate of evaporating, the diffusion coefficient of the solute in the fluid and the driving concentration gradient within the droplet (the difference between dynamic bulk and interfacial concentrations). $U_{c,m}$ and $V_{f,c}$ represent the internal circulation velocity driven by the solutal gradient at zero-field and the electro-solutal advection velocity in the electric field environment. Similar to internal circulation velocity due to thermal gradient, $U_{c,m} = \sigma_c \Delta C_m / \mu$, where $\sigma_c$ is the rate of change of surface tension with respect to change in solutal concentration [14] Substituting the expressions for velocities in eqn. 22, the expression yields

$$\rho \dot{R} R = D \Delta C_m + \frac{\sigma_c (\Delta C_m)^2 R}{\mu} - v(\sqrt{E_{hd}}) \Delta C_m \qquad ()$$

$$\frac{\rho \dot{R} R}{D \Delta C_m} = (1 + Ma_s - Sc\sqrt{E_{hd}}) \qquad ()$$

where $Ma_s$ and $Sc$ are the solutal Marangoni number and Schmidt number. The term $(Sc\sqrt{E_{hd}})$ represents the electro-solutal convection contribution due to the electric field. Rearranging eqn. 24 and putting $Ma_s \gg 1$ [26]

$$\frac{\rho \dot{R} R}{D \Delta C_m} \approx Ma_s - \sqrt{Sc_{sp}} \sqrt{E_{hd} Sc_i} \qquad (26)$$

where $\sqrt{Sc_{sp}}$ represents the species transport Schmidt number and $\sqrt{E_{hd} Sc_i}$ represents the electro-Schmidt number, where $Sc_i$ is the ionic Sc, which is the ratio of mobility (m) to diffusivity of the solvated ions ($Sc_i = m_{ion}/D_{ion}$). The term $(Ma_s - \sqrt{Sc_{sp}} \sqrt{E_{hd} Sc_i})$ behaves as the effective electro-solutal Ma for the system. The electro-Schmidt number essentially indicates the extent of ionic absorption-desorption at the droplet interface due to the applied electric field and its consequences on the internal and interfacial advection dynamics. Figure 12 illustrates the stability plot for the thermal Ma with respect to the solutal Ma ($Ma_T$ and $Ma_s$). Iso-Lewis number lines, proposed by Joo [29], have been used to estimate the stability and behaviour of the internal advection. All the points are found to lie towards the right side of the Le=0 line, with large values of the $Ma_s$ compared to the $Ma_t$, signifying the dominant nature of the solutal Marangoni advection over the thermal Marangoni within the droplet [14, 26]. However, the traditional stability map of the $Ma_t$ and $Ma_s$ is insufficient to determine the effects of electro-solutal effects, as it is observed that hardly any difference in the $Ma_s$ can be observed under field constraints, thereby making the deduction of electro-solutal effects difficult. Figure 13 illustrates the effective thermal Marangoni number ($Ma_{t,e}$) and effective solutal Marangoni number ($Ma_{s,e}$), for variations in field frequency and strength. Similar to fig. 10, two regions have been mapped. In the present case, the field strengths are not potent enough to lead to $\left(Ma_s < Sc\sqrt{E_{HD}}\right)$. It is observed that with increase in the electric field frequency or strength, the points shift towards the origin, indicating reduction in the solutal advection component as well as the thermal component. The direction of the shift with increasing field parameters is evident from the behavior of the iso-$E_{HD}$ lines, where the points shift nearer to the origin with increase in field parameters. However, it is evident that the relative decrease in the value of the $Ma_{s,e}$ is comparatively larger than the reduction in the $Ma_{t,e}$, which showcases that the arrest in the solutal advection by the opposing electro-solutal advection is the dominant mechanism governing the arrest of evaporation. The arrested thermal advection by the opposing electro-thermal advection plays the recessive role. Since the effective electro-solutal Ma is revealed as the dominant governing parameter, the modified circulation velocity under field effect is inversely deduced from Eqn. 22. Representative theoretical velocities are illustrated in fig. 7 (b) and good match is observed with respect to the corresponding spatially averaged circulation velocities. This further cements the finding that the electro-solutal dynamics is the dominant governing mechanism responsible for arrested evaporation.

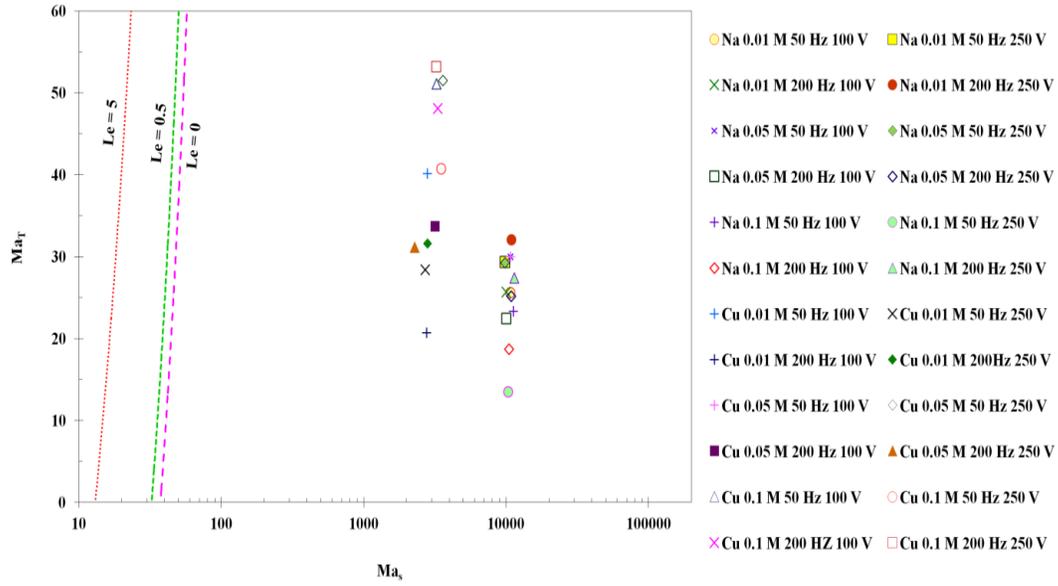

**FIG. 12:** The phase map of the thermal Ma against the solutal Maunder variant field constraints. The iso-Le lines are adapted from literature (Joo [29]).

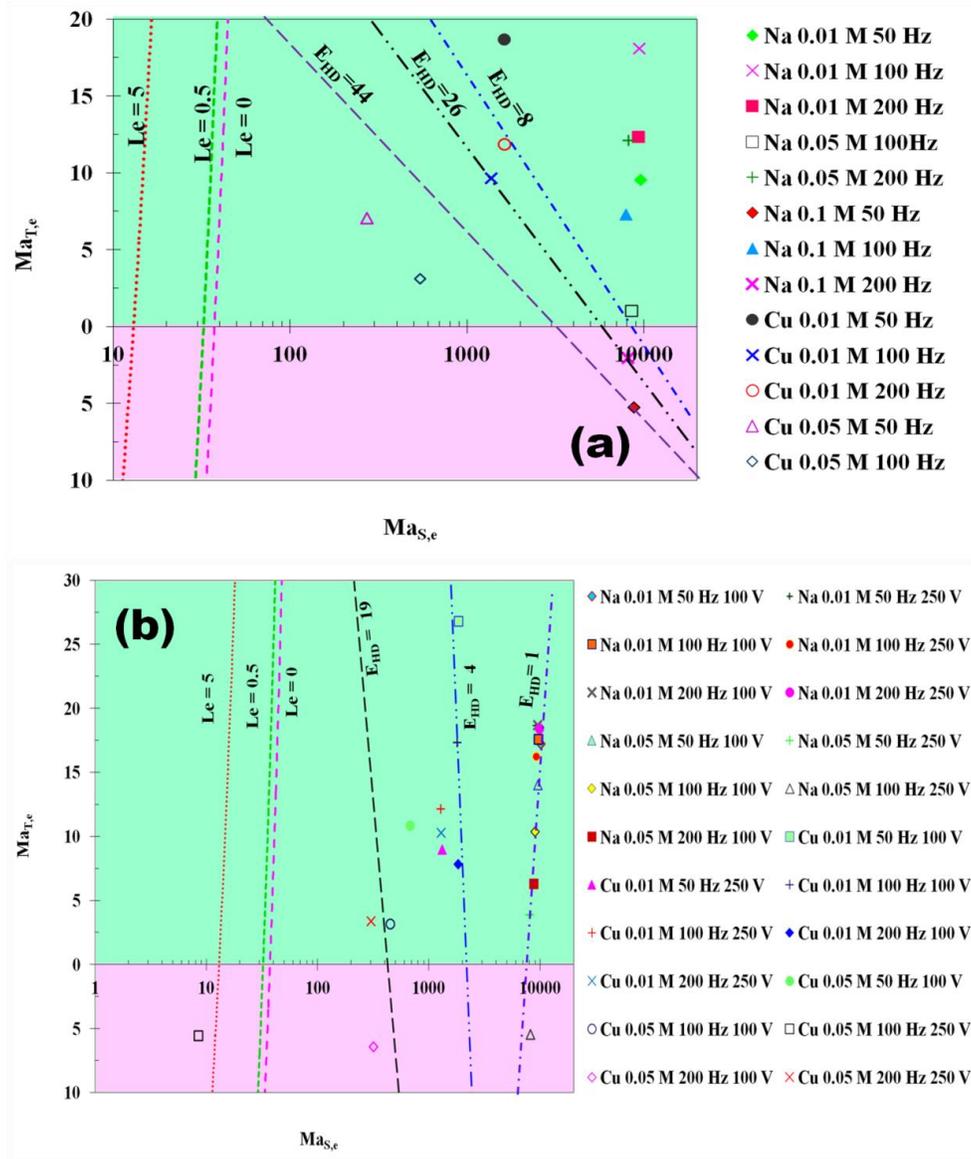

**FIG. 13.** Plot of the effective electro-thermal Ma against the effective electro-solutal Ma. The stability curves are represented by different iso-Le lines (Joo [29]) and iso-$E_{HD}$ lines.

## 4. Conclusions

The present article reports the experimentally observed arrest in evaporation rates of pendant droplets with solvated ions under the influence of an alternating electric field. The evaporation kinetics is observed to conform to the classical $D^2$ law and arrested evaporation from the salt solution droplets is observed. The rate of evaporation is found to deteriorate even below the evaporation rate of water droplet with increase in field strength. The field frequency is observed to improve the evaporation rates, however, it still remains less than the zero-field condition. The same has been explained based on the solvation characterisitcs of the ions. The role of surface tension is tested to determine the root cause, however, it is observed to be insufficient to explain the kinetics. Mathematically, the classical diffusion driven evaporation model is appealed to and it is found to fall short in predicting the arrested evaporation rates. The role of internal circulation is then probed into, and Particle Image Velocimetry reveals atypical arrest in the internal advection velocity within the droplet under influence of electric field. Additionally, change in the orientation of internal circulation dynamics is also observed and has been explained based on electrohydrodynamic considerations. The flow visualization reveals that internal advection dynamics could be a responsible agent for the arrested evaporation rates. A mathematical scaling model is proposed to determine the roles of the thermal and the solutal modes of advection on the internal circulation. The thermal scaling model presents the inability of the thermal Marangoni and Rayleigh convection models to capture the internal advection dynamics. It is deduced that the effective electro-thermal Marangoni convection is a governing parameter, and the electro-diffusive advection (represented by the electro-Prandtl number) is responsible towards arrest of the thermal advection. Further, the role of solutal advection is modelled, and the analysis reveals that the electro-solutal Ma governs the effective advection kinetics. In the solutal advection, the electro-solutal advection (represented by the electro-Schmidt number) opposes the solutal advection (represented by the solutal Ma), leading to arrested internal circulation. Stability mapping of the electro-thermal and electro-solutal effects reveal that with increasing $E_{HD}$, the effective solutal advection decays dominantly compared to the decay of the effective electro-thermal advection. Further, the theoretical circulation velocities derived from the electro-solutal model corroborate well with the PIV observations, thereby establishing the electro-solutal effect as the crux mechanism behind arrested evaporation. The findings on the competitive electrohydrodynamics and electro-solutal effects in ion solvated droplets may find strong implications in microscale electrohydrodynamic thermofluidic transport systems.

## Acknowledgments

PD thanks IIT Ropar for the financial support towards the present research (vide ISIRD grant IITRPR/Research/193). Partial funding from the Department of Mechanical Engineering, IIT Ropar is also thankfully acknowledged.